\documentclass{PoS}

\usepackage{amsmath,amsfonts,amssymb}
\usepackage{graphicx}
\graphicspath{{./figs/}}

\title{Top quark mass effects in Higgs boson pair production up to NNLO}

\ShortTitle{Top quark mass effects in Higgs boson pair production up to
  NNLO}

\author{%
  \speaker{Jens Hoff}\\
  Deutsches Elektronen-Synchrotron~(DESY)\\
  E-mail: \email{jens.hoff@desy.de}}

\abstract{%
  We consider the production of pairs of Standard Model Higgs bosons via
  gluon fusion.  Until recently the full dependence on the top quark
  mass $M_t$ was not known at next-to-leading order.  For this reason we
  apply an approximation based on the expansion for large top quark
  masses up to $\mathcal{O}(1/M_t^{12})$.  At next-to-next-to-leading
  order we avoid the calculation of real corrections via the
  soft-virtual approximation and obtain top quark mass corrections up to
  $\mathcal{O}(1/M_t^4)$.  We use our results to estimate the residual
  uncertainty of the total cross section due to a finite top quark mass
  to be $\mathcal{O}(10\%)$ at next-to-leading order and
  $\mathcal{O}(5\%)$ at next-to-next-to-leading order.}

\FullConference{%
  Loops and Legs in Quantum Field Theory\\
  24-29 April 2016\\
  Leipzig, Germany}

\begin{document}

\section{Introduction}

Higgs boson pair production is the process at the LHC that may in the
future allow for an independent measurement of the cubic Higgs coupling.
With this extraction a test whether the form of the Higgs potential is
consistent with the Standard Model (where the cubic coupling is fixed by
the Higgs bosons' mass $m_H$ and its vacuum expectation value) and
thereby of the mechanism of spontaneous symmetry breaking would be
faciliated.

The dominant production mode is, as for single Higgs boson production
although with a relative suppression of $\mathcal{O}(10^{-3})$, gluon
fusion.  The leading order (LO) calculation was performed retaining the
exact dependence on the top quark mass $M_t$ in
Refs.~\cite{Glover:1987nx,Plehn:1996wb}.  Next-to-leading order (NLO)
and next-to-next-to-leading order (NNLO) corrections were first
calculated in the effective theory where the top quark is integrated
out.  See Ref.~\cite{Dawson:1998py} for the NLO and
Refs.~\cite{deFlorian:2013uza,deFlorian:2013jea} for the NNLO case.
Note that the matching coefficient for Higgs boson pairs differs
starting from three loops from the one for a single Higgs boson, see
Ref.~\cite{Grigo:2014jma}.

Top quark mass corrections at NLO using a systematic expansion in
$1/M_t$ were first studied in
Refs.~\cite{Grigo:2013rya,Grigo:2013xya,Grigo:2014oqa} and in
Ref.~\cite{Grigo:2015dia} this calculation was extended to NNLO.  In
Ref.~\cite{Maltoni:2014eza} the exact dependence on $M_t$ was taken into
account for the real NLO corrections.
Meanwhile the full NLO result became available taking into account the
exact dependence on $M_t$ also for the virtual corrections, see
Ref.~\cite{Borowka:2016ehy}.  For low center-of-mass energies, say
between $\sqrt{s} = 2 m_H$ and $\sqrt{s} = 2 M_t$ the numerical
uncertainties of Ref.~\cite{Borowka:2016ehy} are still quite big whereas
the expansions performed in Refs.~\cite{Grigo:2013rya,Grigo:2015dia}
show a good convergence behaviour.  On the other hand, for higher
center-of-mass energies the results of
Refs.~\cite{Grigo:2013rya,Grigo:2015dia} can only be used to obtain the
order of magnitude of the $M_t$ effects which were estimated to be $\pm
10$ at NLO which is somewhat smaller than the results reported in
Ref.~\cite{Borowka:2016ehy}.

\begin{figure}[t]
  \centering
  \begin{center}
    \includegraphics[scale=0.8]{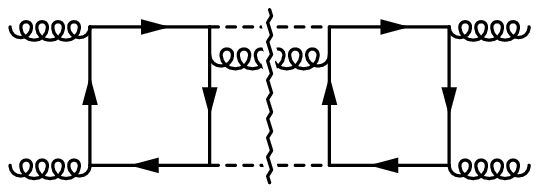}
    \hspace*{1.0pc}
    \includegraphics[scale=0.8]{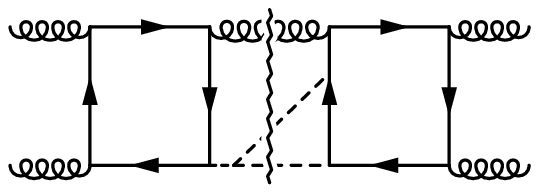}\\
    \bigskip
    \includegraphics[scale=0.8]{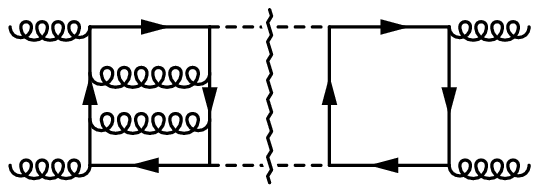}
    \hspace*{1.0pc}
    \includegraphics[scale=0.8]{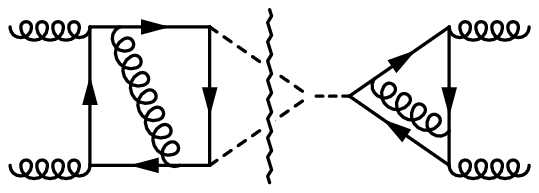}
  \end{center}
  \caption{%
    Sample forward scattering diagrams for the $gg$ channel.  Curly
    lines represent gluons, dashed lines Higgs bosons and solid lines
    top quarks.  The wavy line denotes a cut.  The first row shows real
    corrections at NLO, the second row virtual corrections at NNLO.}
  \label{fig::sample-dias}
\end{figure}

In this contribution we describe the NNLO calculation of
Ref.~\cite{Grigo:2015dia}.  We start with full-theory diagrams where the
top quark has not been integrated out.  We apply the optical theorem on
$gg \to gg$ forward scattering diagrams to extract the imaginary parts
corresponding to real corrections $gg \to HH + X$ with additional
partons $X$ in the final state.  Virtual corrections are calculated
directly from $gg \to HH$ amplitudes by squaring and integration over
the $HH$ phase space.  As a cross check we compute also virtual
corrections via the optical theorem.  In Fig.~\ref{fig::sample-dias} we
show some sample diagrams within the optical theorem approach.  Note, at
NLO (NNLO) we have to consider $gg \to HH$ amplitudes with two (three)
or $gg \to gg$ forward scattering amplitudes with four (five) loops.

\section{Calculation}

\subsection{Differential factorization}

The partonic cross section for the production of a pair of Higgs bosons
via gluon fusion has the perturbative expansion
\begin{align}
  \begin{aligned}
    \sigma_{ij \to HH + X}(s, \rho) &=
    \delta_{ig} \delta_{jg} \sigma_{gg}^{(0)}(s, \rho)
    + \frac{\alpha_s}{\pi}  \sigma_{ij}^{(1)}(s, \rho)
    + \left(\frac{\alpha_s}{\pi}\right)^2  \sigma_{ij}^{(2)}(s, \rho)
    + \ldots\\[0.5pc]
    &= \sigma^{\text{LO}}
    + \delta\sigma^{\text{NLO}}
    + \delta\sigma^{\text{NNLO}}
    + \ldots,
  \end{aligned}
  \label{eqn::part-xsec}
\end{align}
where we consider in the following only the dominant $gg$ channel with
$i = j = g$.  The variable $\rho = m_H^2/M_t^2$ describes the dependence
on the Higgs boson and top quark masses.  For convenience we absorb
powers of $\alpha_s$ in the second line of Eq.~\eqref{eqn::part-xsec}.

The factorization of the LO result can be performed at the level of the
differential cross section, see Ref.~\cite{Dawson:1998py}:
\begin{align}
  \sigma^{(i)} =
  \int_{4 m_H^2}^s \operatorname{d}\!Q^2
  \frac{
    \Big(\frac{
      \operatorname{d}\!\sigma_{\text{exact}}^{(0)}}{
      \operatorname{d}\!Q^2}\Big)}{
    \Big(\frac{
      \operatorname{d}\!\sigma_{\text{exp}}^{(0)}}{
      \operatorname{d}\!Q^2}\Big)}
  \frac{
    \operatorname{d}\!\sigma_{\text{exp}}^{(i)}}{
    \operatorname{d}\!Q^2}
  \quad \text{with} \quad
  \frac{\operatorname{d}\!\sigma_{\text{exp}}^{(i)}}{
    \operatorname{d}\!Q^2} = \sum_{n = 0}^N c_n^{(i)} \rho^n,
  \label{eqn::diff-fact}
\end{align}
where ``exact'' refers to the LO result with full dependence on $\rho$,
``exp'' to the expansion for small $\rho$ and $Q^2$ is the invariant
mass of the Higgs boson pair.  The functional dependence of
$\operatorname{d}\!\sigma_{\text{exact}}^{(0)}/\operatorname{d}\!Q^2$
and $\operatorname{ d}\!\sigma_{\text{exp}}^{(i)}/\operatorname{d}\!Q^2$
in Eq.~\eqref{eqn::diff-fact} are assumed to be similar in the region
where $Q^2 \gtrsim 4 M_t^2$ which is expected to lead to a well behaved
integrand.  Note that we require the series expansions in numerator and
denominator to be truncated at the same order $N$.

Within the framework described in Ref.~\cite{Grigo:2013rya} we computed
the real NLO corrections via the forward scattering amplitude $gg \to
gg$ using the optical theorem.  For this reason we have no immediate
access to the $Q^2$ dependence for these contributions.  In contrast,
the virtual corrections have a trivial $Q^2$ dependence $\delta(s -
Q^2)$ and are available to us from the direct calculation of the~$gg \to
HH$ amplitude.

\subsection{Soft-virtual approximation}

The obstacle we pointed out can be circumvented by applying the
soft-virtual approximation, cf.\ Ref.~\cite{deFlorian:2012za}.  We split
a cross section $\sigma$ up according to
\begin{align}
  \sigma = \text{finite} =
  \sigma^{\text{virt} + \text{ren}}
  + \sigma^{\text{real} + \text{split}} =
  \underbrace{
    \Sigma_{\text{div}} \;+\; \Sigma_{\text{fin}} \quad
    + \quad \Sigma_{\text{soft}}}_{= \Sigma_{\text{SV}}}
  \;+\; \underbrace{\Sigma_{\text{hard}}}_{= \Sigma_{\text{H}}}.
  \label{eqn::sv-approx}
\end{align}
The finite cross section is composed of virtual correction and
renormalization pieces $\sigma^{\text{virt} + \text{ren}}$ and real
correction and infrared counterterm pieces $\sigma^{\text{real} +
  \text{split}}$.  These pieces in turn can be split up further:
divergent terms $\Sigma_{\text{div}}$ and finite terms
$\Sigma_{\text{fin}}$ for the former, divergent ``soft'' terms
$\Sigma_{\text{soft}}$ and finite ``hard'' terms $\Sigma_{\text{hard}} =
\Sigma_{\text{H}}$ for the latter.  The sum of the first three terms on
the right-hand side of Eq.~\eqref{eqn::sv-approx} is finite and
comprises the soft-virtual approximation.  Note that this splitting
holds also for differential cross sections
$\operatorname{d}\!\sigma/\operatorname{d}\!Q^2$.

$\Sigma_{\text{div}}$ is universal for color-less final states and can
be found in Refs.~\cite{Grigo:2014jma,deFlorian:2012za}.  We obtain
$\Sigma_{\text{fin}}$ by computing $\sigma^{\text{virt} + \text{ren}}$
as expansion in $\rho$ and solving $\sigma^{\text{virt} + \text{ren}} =
\Sigma_{\text{div}} + \Sigma_{\text{fin}}$.  $\Sigma_{\text{div}}$ and
$\Sigma_{\text{soft}}$ are proportional to $\sigma^{\text{LO}}$ and
therefore automatically include effects due to finite $M_t$.  We write
the differential and total cross sections as
\begin{align}
  Q^2 \frac{\operatorname{d}\!\sigma}{\operatorname{d}\!Q^2} &=
  \sigma^{\text{LO}} z G(z) \quad
  \text{with} \quad
  G\!\left( z \right) =
  G_{\text{SV}}\!\left( z \right) + G_{\text{H}}\!\left( z \right), \quad
  z = \frac{Q^2}{s},\\
  \sigma &=
  \int_{1 - \delta}^{1} \operatorname{d}\!z
  \sigma^{\text{LO}}(zs) G(z) \quad
  \text{with} \quad \delta = 1 - \frac{4 m_H^2}{s},
\end{align}
where omitting $G_{\text{H}}(z)$ means using the soft-virtual
approximation.  $G_{\text{SV}}(z)$ is constructed from
$\sigma_{\text{fin}}^{(i)}$ and $\sigma^{\text{LO}}$ only and can be
found in Refs.~\cite{Grigo:2015dia,deFlorian:2012za}.

\subsection{Asymptotic expansion}

\begin{figure}[t]
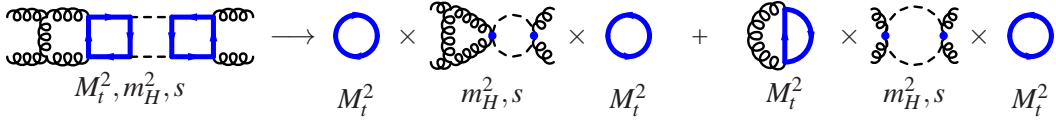

  \centering
  \raisebox{-9pt}{\parbox{100pt}{
      \centering
      \includegraphics{NLO_asy-exp-2.1}\\
      $M_t^2, m_H^2, s$}}
  $\longrightarrow$
  \raisebox{-8.5pt}{\parbox{25pt}{
      \centering
      \includegraphics{NLO_asy-exp-2.2}\\
      $M_t^2$}}
  $\times$
  \raisebox{-8.5pt}{\parbox{50pt}{
      \centering
      \includegraphics{NLO_asy-exp-2.3}\\
      $m_H^2, s$}}
  $\times$
  \raisebox{-8.5pt}{\parbox{25pt}{
      \centering
      \includegraphics{NLO_asy-exp-2.4}\\
      $M_t^2$}}
  \hspace*{0.5pc} + \hspace*{0.5pc}
  \raisebox{-8.5pt}{\parbox{35pt}{
      \centering
      \includegraphics{NLO_asy-exp-2.5}\\
      $M_t^2$}}
  $\times$
  \raisebox{-8.5pt}{\parbox{35pt}{
      \centering
      \includegraphics{NLO_asy-exp-2.6}\\
      $m_H^2, s$}}
  $\times$
  \raisebox{-8.5pt}{\parbox{25pt}{
      \centering
      \includegraphics{NLO_asy-exp-2.7}\\
      $M_t^2$}}
  \caption{%
    Asymptotic expansion in $M_t^2 \gg m_H^2, s$ applied to a virtual
    NLO forward scattering diagram, resulting in two different regions.
    Curly lines are gluons, dashed lines are the (cut) Higgs bosons and
    thick blue lines represent the top quarks.  Each (sub)diagram is
    labeled with the scales it involves.}
  \label{fig::asy-exp}
\end{figure}

Let us briefly describe the computation of the diagrams as an expansion
in $\rho$.  The integrands of Feynman integrals are expanded according
to a hierarchy of scales $M_t^2 \gg m_H^2, s$ for all possible scalings
of loop momenta, so-called ``regions'', and summed afterwards.  The
outcome of this procedure is a reduction of scales and loops which have
to be considered at the same time (diagrams factorize).  In case of an
expansion for a hard mass all relevant regions correspond to subgraphs
which must contain all heavy lines.  For illustration we sketch the
expansion regions for a virtual NLO diagram in the forward scattering
approach in Fig.~\ref{fig::asy-exp}.  Two regions emerge: one with a
``soft'' two-loop four-point graph multiplied with two ``hard'' one-loop
tadpoles and one with a soft one-loop four-point graph multiplied with
hard one- and two-loop tadpoles.

\subsection{Software setup}

Our software setup is highly automated, but we omit a detailed survey
here and refer instead to Ref.~\cite{Grigo:2015dia} where also
intermediate results are given and the calculation of the master
integrals is discussed.  We generate diagrams with {\tt
  QGRAF}~\cite{Nogueira:1991ex} where in the case of $gg \to gg$
postprocessing~\cite{Grigo:2014oqa,diss_Hoff} is mandatory.  For
topology identification and other steps of the calculation we use the
package {\tt TopoID}~\cite{Grigo:2014oqa,diss_Hoff}.  Asymptotic
expansion and mapping of diagrams to topologies is performed with {\tt
  q2e} and {\tt exp}~\cite{Harlander:1997zb,Seidensticker:1999bb}.  The
reduction to scalar integrals uses {\tt FORM}~\cite{Kuipers:2012rf}.
Soft four-point subdiagrams are reduced to master integrals with {\tt
  FIRE}~\cite{Smirnov:2013dia,Smirnov:2014hma} and the in-house code
{\tt rows}~\cite{diss_Hoff}.  Hard subdiagrams are always massive
tadpoles and can be treated with {\tt MATAD}~\cite{Steinhauser:2000ry}.

\section{Results}

We summarize the main features of our findings in bullet points.  For
details, such as the particular values of input parameters, cf.\
Ref.~\cite{Grigo:2015dia}.  Throughout the presented analysis we set the
renormalization scale to $\mu = 2 m_H$ and use the MSTW2008
PDFs~\cite{Martin:2009iq}.

\begin{itemize}

\item In a split-up (not shown) of the NLO correction to the total
  partonic cross section into soft-virtual and hard contributions, we
  observe different patterns when including higher $\rho$ corrections:
  soft-virtual corrections increase, whereas hard ones descrease with
  $\sqrt{s}$.  Soft-virtual corrections dominate over the full range of
  $\sqrt{s}$, above $400\,\text{GeV}$ hard ones become flat.

\item In Fig.~\ref{fig::NLO-LO-had-fac} we show results for hadronic
  quantities.  We introduced a technical upper cut-off for the partonic
  center-of-mass energy $\sqrt{s_{\text{cut}}}$ which is a good
  approximation to the invariant mass of the produced Higgs boson pair:
  \begin{align}
    \sigma_H(s_H, s_{\text{cut}}) =
    \int_{4 m_H^2/s_H}^1 \operatorname{d}\!\tau
    \left(\frac{\operatorname{d}\mathcal{L}_{gg}}{
        \operatorname{d}\!\tau}\right)\!(\tau)
    \sigma(\tau s_H) \theta(s_{\text{cut}} - \tau S_H),
  \end{align}
  where $\sqrt{s_H} = 14\,\text{TeV}$ is the hadronic center-of-mass
  energy for the LHC and $\mathcal{L}_{gg}$ is the luminosity function
  for two gluons in the inital state.

  From the spread of $\rho$ orders for the total hadronic cross section
  $\sigma_H^{\text{NLO}}$ on the right-hand side, when
  $\sqrt{s_{\text{cut}}} \to \infty$, we infer the uncertainty due to
  top quark mass corrections to be about~$\pm10\%$.

  \begin{figure}[t]
    \centering
    \includegraphics[scale=0.8]{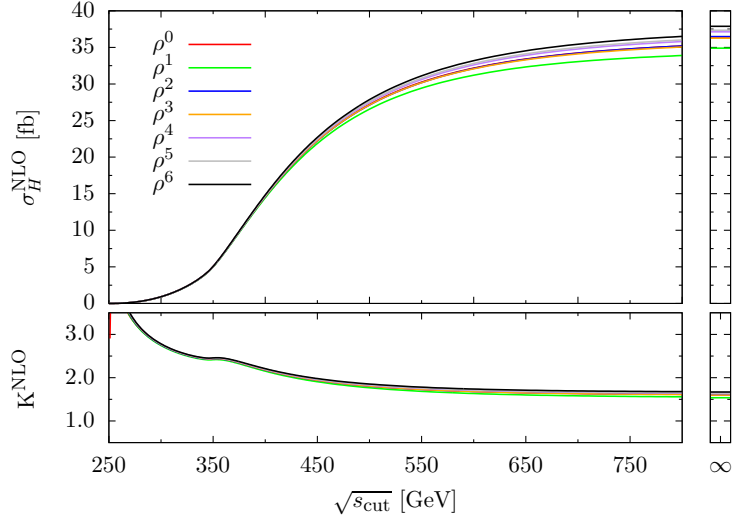}
    \caption{%
      NLO hadronic cross section $\sigma_H^{\text{NLO}}$ in the upper
      panel and $K$ factor $K^{\text{NLO}}$ in the lower panel as
      functions of $\sqrt{s_{\text{cut}}}$, a technical upper cut on
      $\sqrt{s}$ and proxy to the invariant mass of the Higgs boson
      pair.  We use ``$\infty$'' to symbolize the results for the total
      inclusive cross section and $K$ factor on the right-hand side.
      Here and in the following the color coding indicates the inclusion
      of higher orders in the $\rho$ expansion.  Figure taken from
      Ref.~\cite{Grigo:2015dia}.}
    \label{fig::NLO-LO-had-fac}
  \end{figure}

\item In the soft-virtual approximation $G_{\text{SV}}(z)$ from
  Eq.~\eqref{eqn::sv-approx} can be replaced by $f(z) G_{\text{SV}}(z)$
  with any $f(z)$ fulfilling $f(1) = 1$ since the splitting into hard
  and soft-virtual components is not unique.  At NLO we observe that
  using $f(z) = z$ and neglecting hard contributions is accurate within
  $2\%$.  Also, replacing $\log(\mu^2/s)$ by $\log(\mu^2/Q^2)$ leads to
  better results which can be justified in the soft limit where $s
  \approx Q^2$.  We adopt these prescriptions to proceed at NNLO.

\item In Fig.~\ref{fig::NNLO-part-0-z} we recognize for the LO, NLO and
  NNLO corrections $\delta\sigma$ the same pattern in the $\rho$
  expansion (negative shifts for $\rho^1$ and positive ones for
  $\rho^2$) and that the peak positions move to lower values for
  $\sqrt{s}$ for higher perturbative orders.

  \begin{figure}[t]
    \centering
    \includegraphics[scale=0.8]{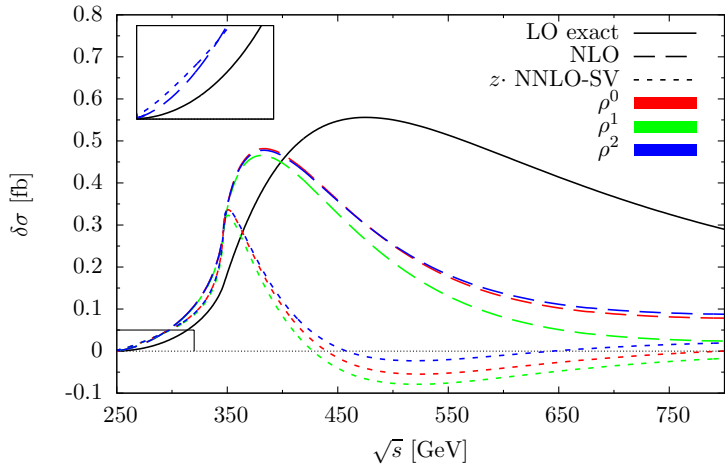}
    \caption{%
      LO, NLO and NNLO contributions $\delta\sigma$ to the partonic
      cross section.  At LO the exact result is shown as solid black
      line, at NLO and NNLO we give only the first three expansion
      orders in $\rho$ for consistency and we use $f(z) = z$ at NNLO
      (see the main text).  The inset magnifies the region of small
      $\sqrt{s}$.  Figure taken from Ref.~\cite{Grigo:2015dia}.}
    \label{fig::NNLO-part-0-z}
  \end{figure}

\item For the total hadronic cross section $\sigma_H$ up to NNLO in
  Fig.~\ref{fig::NNLO-had-0-z} we find good convergence up to
  $\sqrt{s_{\text{cut}}} \approx 400\,\text{GeV}$ and deduce in the same
  way as on NLO an uncertainty due to the top quark mass of about
  $\pm5\%$ (note that NNLO corrections within the effective theory
  amount to about $20\%$ by themselves).

  \begin{figure}[t]
    \centering
    \includegraphics[scale=0.8]{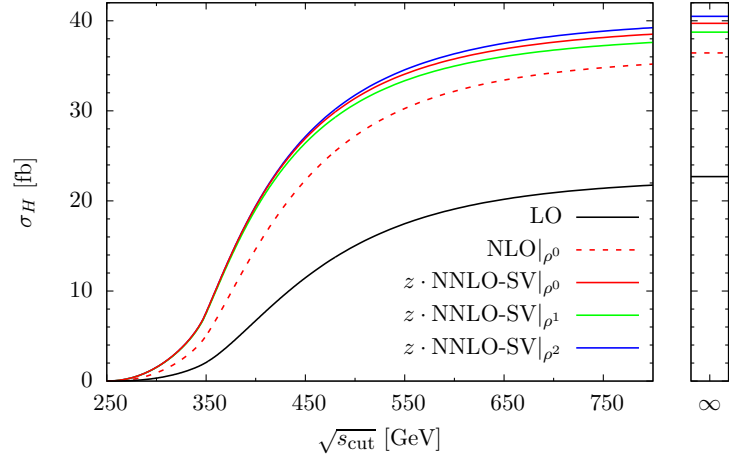}
    \caption{%
      LO, NLO and NNLO hadronic cross sections $\sigma_H$.  At LO the
      exact is shown, at NLO we give only the leading expansion term and
      at NNLO the first three terms in $\rho$.  On the right-hand side
      the total inclusive results are given.  Figure taken from
      Ref.~\cite{Grigo:2015dia}.}
    \label{fig::NNLO-had-0-z}
  \end{figure}

\item In the behavior of the $K$ factor up to NNLO in
  Fig.~\ref{fig::NNLO-had-K-0-z} we see that the characteristic form
  around the $2 M_t$ threshold is not washed out.  The strong raise
  close to the $2 m_H$ threshold is explained by the steepness of the
  NNLO correction, see the inset.  The hadronic NNLO $K$~factor is in
  the range $1.7$ to $1.8$.

  \begin{figure}[t]
    \centering
    \includegraphics[scale=0.8]{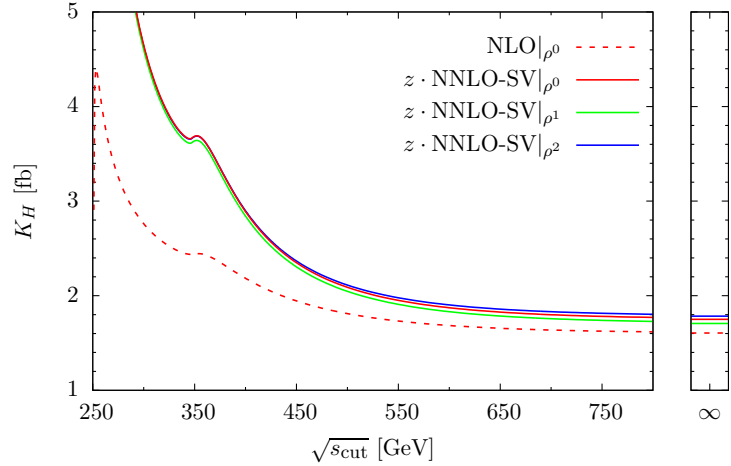}
    \caption{%
      LO, NLO and NNLO hadronic $K$ factors $K_H$.  The notation is as
      in Fig.~\protect\ref{fig::NNLO-had-0-z}.  Figure taken from
      Ref.~\cite{Grigo:2015dia}.}
    \label{fig::NNLO-had-K-0-z}
  \end{figure}

\end{itemize}

\section{Conclusion}

We computed corrections due to a finite top quark mass using an
asymptotic expansion in the limit $M_t^2 \gg m_H^2, s$.  At NLO our
method yields results up to $\mathcal{O}(1/M_t^{12})$, at NNLO up to
$\mathcal{O}(1/M_t^{4})$ using the soft-virtual approximation.  We
estimate the residual error on the total cross section due to finite
$M_t$ to be $\mathcal{O}(10\%)$ at NLO and $\mathcal{O}(5\%)$ at NNLO.

The recently completed full NLO contribution to the total cross section,
see Ref.~\cite{Borowka:2016ehy} and the
presentations~\cite{talk_kerner,talk_jones}, is decreased by $14\%$
compared to the $M_t \to \infty$ limit.  For $Q^2 \le 400\,\text{GeV}$
effects of $\mathcal{O}(10\%)$ are reported for the differential cross
section and even larger ones above $400\,\text{GeV}$.

The NNLO contributions yield a $\mathcal{O}(20\%)$ correction in the
$M_t \to \infty$ limit which could be modified substantially by the
$M_t$ dependence, but a full NNLO calculation is out of scope of present
techniques.  Therefore it seems disirable to refine our approximation
procedure to better reproduce the findings of
Ref.~\cite{Borowka:2016ehy} and to revisit the NNLO case.

\end{document}